\documentclass[aps,prb,twocolumn,showpacs,superscriptaddress]{revtex4-1}

\usepackage{graphicx}
\usepackage{multirow}
\usepackage{color}

\begin{document}


\title{Magnetic instability induced by Rh-doping in Kondo semiconductor CeRu$_2$Al$_{10}$}


\author{Hanjie Guo}
\email[]{hjguo@zju.edu.cn}
\affiliation{Department of Physics and State Key Laboratory of Silicon Materials, Zhejiang University, Hangzhou 310027, China}
\affiliation{Advanced Meson Science Laboratory, RIKEN, 2-1 Hirosawa, Wako, Saitama 351-0198, Japan}

\author{Hiroshi Tanida}
\affiliation{Department of Quantum Matter, ADSM, Hiroshima University, Higashihiroshima, Hiroshima 739-8530, Japan}

\author{Riki Kobayashi}
\affiliation{Quantum Condensed Matter Division, Oak Ridge National Laboratory, Oak Ridge, Tennessee 37831, USA}
\affiliation{Neutron Science Laboratory, Institute for Solid State Physics, University of Tokyo, Tokai, 319-1106, Japan}

\author{Ikuto Kawasaki}
\affiliation{Advanced Meson Science Laboratory, RIKEN, 2-1 Hirosawa, Wako, Saitama 351-0198, Japan}

\author{Masafumi Sera}
\affiliation{Department of Quantum Matter, ADSM, Hiroshima University, Higashihiroshima, Hiroshima 739-8530, Japan}

\author{Takashi Nishioka}
\affiliation{Graduate School of Integrated Arts-and-Science, Kochi University, Kochi 780-8520, Japan}

\author{Masahiro Matsumura}
\affiliation{Graduate School of Integrated Arts-and-Science, Kochi University, Kochi 780-8520, Japan}

\author{Isao Watanabe}
\affiliation{Advanced Meson Science Laboratory, RIKEN, 2-1 Hirosawa, Wako, Saitama 351-0198, Japan}

\author{Zhu-an Xu}
\affiliation{Department of Physics and State Key Laboratory of Silicon Materials, Zhejiang University, Hangzhou 310027, China}


\date{\today}

\begin{abstract}
Magnetic ground state of Rh-doped Kondo semiconductor CeRu$_2$Al$_{10}$ [Ce(Ru$_{1-x}$Rh$_x$)$_2$Al$_{10}$] is investigated by muon-spin relaxation method. Muon-spin precession with two frequencies is observed in the $x$ = 0 sample, while only one frequency is present in the $x$ = 0.05 and 0.1 samples, which is attributed to the broad static field distribution at the muon site. The internal field at the muon site is enhanced from about 180 G in $x$ = 0 sample to about 800 G in the Rh-doped samples, supporting the spin-flop transition as suggested by macroscopic measurements, and the boundary of different magnetic ground states is identified around $x$ = 0.03. The drastic change of magnetic ground state by a small amount of Rh-doping (3\%) indicates that the magnetic structure in CeRu$_2$Al$_{10}$ is not robust and can be easily tuned by external perturbations such as electron doping. The anomalous temperature dependence of internal field in CeRu$_2$Al$_{10}$ is suggested to be attributed to the hyperfine interaction between muons and conduction electrons.
\end{abstract}


\maketitle

\section{Introduction}
Ternary compounds Ce\textit{T}$_2$Al$_{10}$ (\textit{T} = Ru, Os and Fe), which are characterized as Kondo semiconductors, have attracted much attention due to their unusual magnetic properties.\cite{Nishioka_novel_phase, Strydom_thermo, Tanida_NMR, Tanida_magnetic_anisotropy, Kondo_high_field_Ru, Kondo_high_field_Os, Matsumura_NQR, Matsumura_reconciliation, Muro_Fe_gap, Strigari_crystal_field, Robert_excitations, Takashi_Ru_Fe, Tanida_fine_structure, Robert_Ru_spin_dynamics} Magnetic ordering is absent in CeFe$_2$Al$_{10}$ down to 40 mK,\cite{Adroja_muSR} while CeRu$_2$Al$_{10}$ and CeOs$_2$Al$_{10}$ are antiferromagnetically ordered with magnetic moment, $m_\mathrm{AF}$ along the \textit{c}-axis below about 27 K and 29 K, respectively.\cite{Kato_neutron_single_crystal} The transition temperature, $T_0$ is too high considering that the nearest-neighbor distance between two Ce ions is about 5.2 {\AA}.\cite{Nishioka_novel_phase} Magnetic susceptibility measurement shows large anisotropy ($\chi_a > \chi_c > \chi_b$) above $T_0$, where $\chi_i$ is the magnetic susceptibility along the $i$ direction. The easy axis is the \textit{a}-axis while the $m_\mathrm{AF}$ $\parallel$ \textit{c} with $m_\mathrm{AF}$ = 0.42 and 0.29 $\mu_\mathrm{B}$/Ce for $T$ = Ru and Os, respectively.\cite{Kato_neutron_single_crystal} Strigari \textit{et al}. have shown that the crystal electric field (CEF) scheme obtained from soft X-ray absorption spectroscopy can explain the easy axis and the small magnetic moment on CeRu$_2$Al$_{10}$.\cite{Strigari_crystal_field} Kunimori \textit{et al}. also performed mean-field calculations and showed that the CEF scheme only is not enough to account for the magnetic order below $T_0$ and pointed out the importance of the conduction- and \textit{f}-electron (\textit{c-f}) hybridization for the unusual magnetic ordered state.\cite{Kunimori_cf} Kondo \textit{et al}. speculated that the strong \textit{c-f} hybridization suppresses the spin degrees of freedom along the \textit{a}-axis and forces the $m_\mathrm{AF}$ to be along the \textit{c}-axis in place of the \textit{a}-axis.\cite{Kondo_high_field_Os} By applying a magnetic field along the \textit{c}-axis, a spin-flop transition is observed from $m_\mathrm{AF} \parallel c$  to $m_\mathrm{AF} \parallel b$ when the magnetic field is beyond the characteristic value $H^*$ of about 4 T, even the $\chi_b$ is the smallest in the paramagnetic state.\cite{Tanida_NMR} Despite much effort to elucidate the magnetic properties of this system, the origin of the magnetic ordering is still ambiguous.

Muon-spin relaxation ($\mu$SR) experiments have been performed on both CeRu$_2$Al$_{10}$ and CeOs$_2$Al$_{10}$ to investigate the unusual magnetic properties on this system.\cite{Kambe_muSR, Khalyavin_muSR, Adroja_muSR} $\mu$SR is a very sensitive local probe of small magnetic fields due to the large moment of muon. Due to the local probe character, volume fraction of different phases or muon positions can be obtained by analyzing the amplitude of each component. The muon-spin implanted into the sample will precess about the transverse local field if the internal field is not canceled out at the muon site in the magnetic ordered state, and the muon-spin polarization will not be relaxed if the internal field is in parallel with the initial muon-spin direction. In a polycrystalline sample of fully magnetic ordered state, the ratio of 1 : 3 is expected between the non-precession and the total muons since one third of the muons will sense the internal field along the initial muon-spin direction.

First direct evidence of the magnetic ground state on CeRu$_2$Al$_{10}$ was unraveled by zero field (ZF) $\mu$SR experiment on the single crystal performed by Kambe \textit{et al.}\cite{Kambe_muSR} Another $\mu$SR experiment was also performed on the polycrystalline sample by improving the data statistics and two muon frequencies were observed, suggesting at least two muon sites appear in the sample.\cite{Khalyavin_muSR} Both results exhibit a decrease of internal field at the muon site below about 20 K. However, the origin of the decrease is not well understood. $\mu$SR experiment on CeOs$_2$Al$_{10}$ was carried out by Adroja \textit{et al.} and three frequencies were identified.\cite{Adroja_muSR} One of these components also exhibits unusual behaviors, showing a dip at about 11 K in the temperature dependence, and was suggested to be probably related to a structure distortion or resistivity anomaly.

Recently, Kondo \textit{et al.} reported the macroscopic measurement on Ce(Ru$_{0.95}$Rh$_{0.05}$)$_2$Al$_{10}$ using high magnetic field and showed that the magnetic susceptibility is more Curie-Weiss like at high temperatures compared with the undoped sample and decreases drastically below $T_0$ when the magnetic field is applied along the \textit{a}-axis, indicating that the \textit{c-f} hybridization becomes weaker and the 4\textit{f} electrons are more localized. The occurrence of a spin-flop transition is observed when a magnetic field larger than about 13 T is applied along the \textit{a}-axis, suggesting that the $m_\mathrm{AF}$ is along the \textit{a}-axis in the Rh-doped sample.\cite{Kondo_Rh_doped} Specific heat measurement on the Rh-doped samples show that $T_0$ is suppressed with the increase of Rh-content and $T_0$ is about 23 K for the Ce(Ru$_{0.88}$Rh$_{0.12}$)$_2$Al$_{10}$ sample.\cite{Kobayashi_Rh_transport} The Rh has one more electron than Ru and even a small amount of Rh-doping induces significant change in the magnetic properties. In order to shed new light on the influence of the Rh-doping effect, we investigate the magnetic properties from a microscopic way using $\mu$SR method on a series of Ce(Ru$_{1-x}$Rh$_x$)$_2$Al$_{10}$ with $x$ = 0, 0.03, 0.05 and 0.1.

In this paper, we present that the internal fields at the muon sites are changed drastically from $x$ = 0 to $x$ = 0.1. Muon-spin precession with two frequencies is observed on the $x$ = 0 sample, while only one frequency is present in the $x$ = 0.05 and 0.1 samples, which is attributed to the broad static field distribution at the muon site. The internal fields in the Rh-doped samples follow a universal curve and clearly deviate from the mean-field-like behavior, and are about four times larger than that of the $x$ = 0 sample at the base temperature, 2.7 K. Compared with the dipolar field calculation, the enhancement of internal field at the muon site supports the spin-flop transition suggested by the macroscopic measurement. We identify the boundary of different magnetic ground states around $x$ = 0.03, suggesting that the magnetic ground state in CeRu$_2$Al$_{10}$ is not robust and can be tuned easily by chemical doping. By comparing with previous results from neutron scattering, thermoelectric power, NQR and $\mu$SR experiments, we also propose that the hyperfine interaction between muon and conduction electrons may induce the anomalous temperature dependence of the internal field in CeRu$_2$Al$_{10}$.

\section{Experiment}
Single crystals of Ce(Ru$_{1-x}$Rh$_x$)$_2$Al$_{10}$ ($x$ = 0, 0.03, 0.05 and 0.1) were synthesized by the Al self-flux method as reported.\cite{Tanida_pressure_resistivity} $\mu$SR experiments were performed at the RIKEN-RAL muon facility at Rutherford-Appleton Laboratory, U.K.\cite{Matsuzaki_Facility} Many pieces of single crystals were packed randomly using high-purity silver foil, therefore, the obtained composite was regarded as a polycrystal. The spin-polarized muons were injected into the sample, and the decayed positrons ejected preferentially along the muon-spin direction were accumulated at various temperatures. The typical muon events are about 20 million at each measurement. The initial muon-spin polarization is in parallel with the beam line. Forward and backward counters are located in the upstream and downstream of the beam, respectively. The asymmetry parameter of the muon-spin polarization is defined as $A(t)=[F(t)-\alpha B(t)]/[F(t)+ \alpha B(t)]$, where $F(t$) and $B(t)$ are the number of muon events counted by the forward and backward counters at time $t$, respectively. $\alpha$ reflects the relative counting efficiency of forward and backward counters. The experiments were carried out in ZF and longitudinal field (LF) configurations. The LF was applied along the initial muon-spin polarization.

\section{experimental result}

\begin{figure}
\includegraphics[width=7cm]{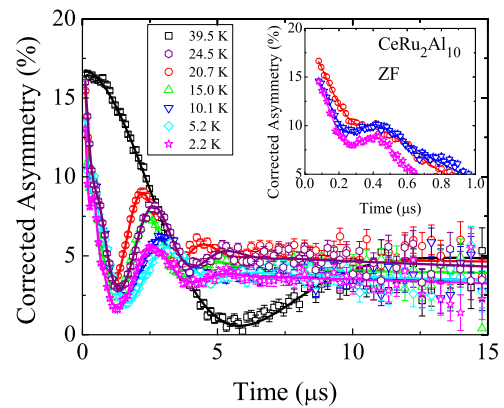}
\caption{(Color online) Zero-field time spectra for CeRu$_2$Al$_{10}$ at various temperatures. The inset in the early time scale shows the fast muon-spin precession component. Solid curves are fitting results to the Eq. (1) and (2).\label{undoped_spectrum}}
\end{figure}

Figure \ref{undoped_spectrum} shows the time dependence of asymmetry ($\mu$SR spectra) on CeRu$_2$Al$_{10}$ measured at various temperatures in ZF. The results are roughly consistent with previous results reported by other groups.\cite{Kambe_muSR, Khalyavin_muSR} The spectra show Kubo-Toyabe behavior above $T_0$ originated from the nuclear moments of Ru and Al. As the temperature is cooled down below $T_0$, muon-spin precession with two frequencies is observed, indicating that at least two muon sites exist in the compound. The inset of Fig. \ref{undoped_spectrum} plotted in an early time scale shows the fast-precession component.

\begin{figure}
\includegraphics[width=8.6cm]{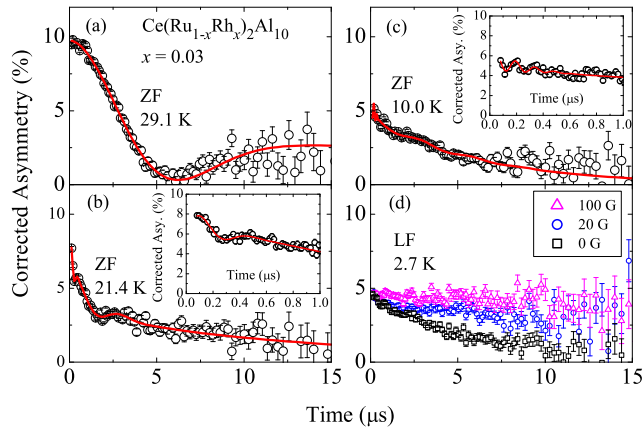}
\caption{(Color online) Time spectra at various temperatures for $x = 0.03$ compound. (a)-(c) show that in ZF condition, the insets show the spectra in an early time region. (d) shows the LF measurement at 2.7 K. The red curves are fitted to Eq. (1) for (a) and to Eq. (2) for (b) and (c).\label{x_003_spectrum}}
\end{figure}

Figures \ref{x_003_spectrum}(a)-\ref{x_003_spectrum}(d) plot the $\mu$SR spectra at various temperatures on the $x$ = 0.03 sample. Kubo-Toyabe behavior is observed above $T_0$ in ZF, as shown in Fig. \ref{x_003_spectrum}(a). With decreasing temperature below $T_0$, muon-spin precession with two frequencies is observed, as shown in Fig. \ref{x_003_spectrum}(b) and \ref{x_003_spectrum}(c). A loss of initial asymmetry at $t$ = 0, $A(0)$, is also observed due to limiting response of the pulsed muon source to the large internal field. With further decrease in temperature, muon-spin precession is not observable below $\sim$ 8 K, as shown in Fig. \ref{x_003_spectrum}(d). By applying a small LF of 100 G, the long time asymmetry is recovered, and the spectrum becomes flat, while the initial asymmetry is not changed.

\begin{figure}
\includegraphics[width=8.6cm]{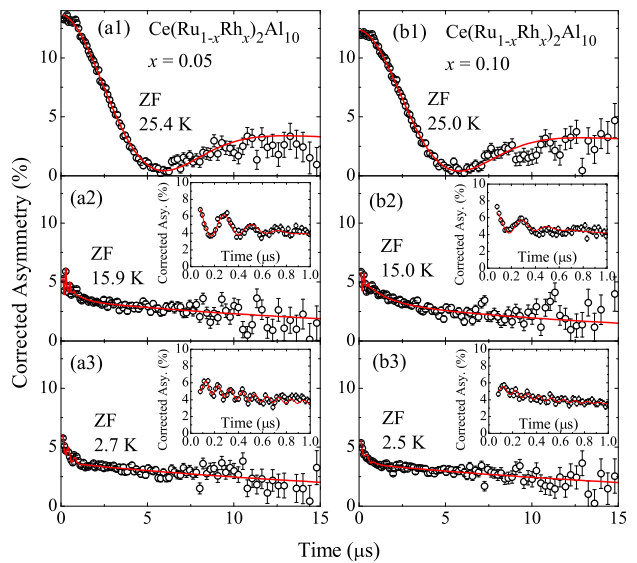}
\caption{(Color online) Time spectra at various temperatures for $x = 0.05$ (left) and $x = 0.1$ (right). The insets in the early time scale show the fast muon-spin precession. Solid curves are fitting results to Eq. (1) and (2).\label{x_005_010_spectrum}}
\end{figure}

Figures \ref{x_005_010_spectrum}(a1)-\ref{x_005_010_spectrum}(b3) show the time spectra in ZF for the $x$ = 0.05 and 0.1 samples. The Kubo-Toyabe behavior is observed above $T_0$, while a loss of initial asymmetry is observed at lower temperatures, and the early time region as shown in the insets show muon-spin precessions in both samples down to the base temperature. However, only a fast precession frequency is observed, while the other component is absent compared with the $x$ = 0 and 0.03 samples. The spectra in these two compounds exhibit similar behaviors, suggesting the same ground state.

\begin{figure}
\includegraphics[width=8.6cm]{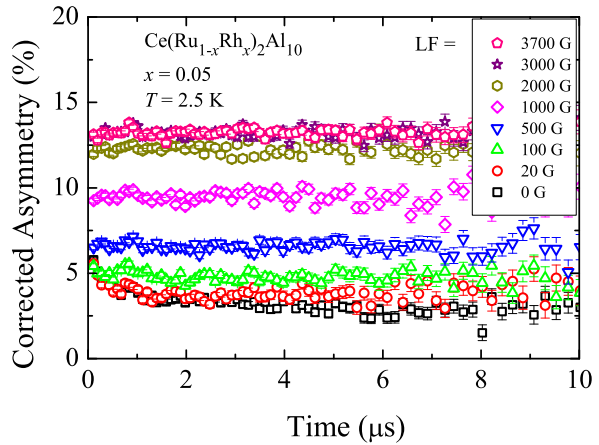}
\caption{(Color online) LF measurements at the base temperature on the $x$ = 0.05 sample. \label{x_005_LF}}
\end{figure}

LF measurement at the base temperature on the $x$ = 0.05 samples is shown in Fig. \ref{x_005_LF}. When the LF is less than 100 G, the long time asymmetry is recovered with the increase of LF, while the initial asymmetry is not affected, similar to the result on the $x$ = 0.03 sample. In addition, the spectrum in LF = 100 G is nearly flat, such a behavior strongly suggests that the internal field is static in the time window of $\mu$SR. When LF $>$ 100 G, the spectra keep flat while the initial asymmetry is recovered gradually, and the full asymmetry is achieved when LF is larger than 3000 G.

In order to analyze these spectra, we use the function of
\begin{equation}\label{eq_Above_T0}
    A(t)=A(0)G_{\rm KT}(t)\mathrm{exp}(-\lambda t)
\end{equation}
above $T_0$, where
\begin{equation}\label{Kubo-Toyabe}
    G_{\rm KT}(t)=\frac{1}{3}+\frac{2}{3}(1-\Delta^2 t^2)\mathrm{exp}(-\frac{\Delta^2 t^2}{2})
\end{equation}
is the Kubo-Toyabe function,\cite{Hayano_Kubo_Toyabe_function} usually used to describe the $\mu$SR spectrum in the randomly oriented static internal fields with Gaussian distribution, which originates from the nuclear dipole moments of Ru ($^{99}$Ru, $I$ = 3/2, 12.72\% abundance; $^{101}$Ru, $I$ = 5/2, 17.07\% abundance), $^{27}$Al ($I$ = 5/2) and $^{103}$Rh ($I$ = 1/2) in this case. $A(0)$ is the initial asymmetry, $\Delta/\gamma_\mu$ represents the distribution width of the local fields, $\gamma_\mu/2\pi$ = 13.55 MHz/kOe is the gyromagnetic ratio of muon, and $\lambda$ is the depolarization rate caused by fluctuating electronic spins. In all these analyses, $\Delta$ is about 0.3 $\mu \mathrm{s}^{-1}$ and temperature independent above $T_0$, $\lambda$ is negligible at high temperatures and shows an increase when approaching $T_0$ from above. The background component, $A_\mathrm{bag}$ originating from the muons stopped at the silver sample holder, is subtracted from the spectra by the fitting result at high temperatures and fixed for all the other analyses.

Below $T_0$, the spectra are described uniformly by the function
\begin{equation}\label{}
    A(t)=\sum_i A_{i}\mathrm{cos}(\gamma_{\mu}H_i t+\phi_i)\mathrm{e}^{-\sigma_i^2 t^2}+\sum_j A_{j}\mathrm{e}^{-\lambda_{j} t},
\end{equation}
where $H_i$, $i=\rm{\{small, large\}}$, is the internal field at the muon site, $\sigma_i$ and $\phi_i$ are the damping rate and phase of the muon-spin precession, respectively. $\lambda_j$, $j=\mathrm{\{slow, fast\}}$, is muon-spin relaxation rate. $A_i$ and $A_j$ are the amplitudes of each component. In the case that only the fast muon-spin precession appears, $A_\mathrm{small}$ is set to be zero. The fitted results are the red curves in Fig's. 1-3.

\begin{figure}
\includegraphics[width=7cm]{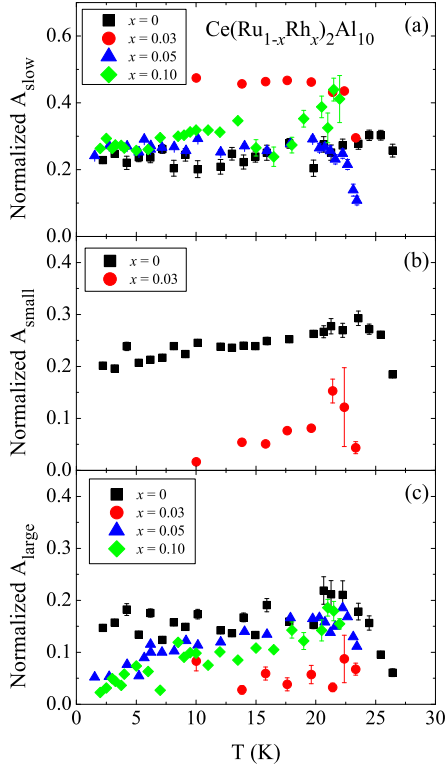}
\caption{(Color online) Temperature dependence of the normalized amplitudes of (a) $A_\mathrm{slow}$, (b) $A_\mathrm{small}$ and (c) $A_\mathrm{large}$.\label{normalized_amplitude}}
\end{figure}

Figures \ref{normalized_amplitude}(a)-\ref{normalized_amplitude}(c) show the temperature dependence of the normalized amplitudes of $A_\mathrm{slow}$, $A_\mathrm{small}$ and $A_\mathrm{large}$, respectively. As shown in Fig. \ref{normalized_amplitude}(a), $A_\mathrm{slow}$ in the $x$ = 0, 0.05 and 0.1 samples are comparable to each other and about 0.25 at the base temperature. While for the $x$ = 0.03 sample, $A_\mathrm{slow}$ is much higher and of the value about 0.5 in the measured temperature range. As the $A_\mathrm{slow}$ component should also reflect the 1/3 component of the muons with their initial spin direction parallel to the internal field, and the values for the $x$ = 0, 0.05 and 0.1 are close to the ideal one, the magnetic volume fraction in these samples can be considered to be nearly full. Figure \ref{normalized_amplitude}(b) shows the temperature dependence of the oscillation amplitude $A_\mathrm{small}$. For the $x$ = 0 sample, it is about 0.3 at 20 K but gradually decreases with decreasing temperature, and reaches the value of about 0.2 at the base temperature. $A_\mathrm{small}$ in the $x$ = 0.03 sample is much smaller compared with the $x$ = 0 sample, and disappears below about 10 K. Figure \ref{normalized_amplitude}(c) shows the temperature dependence of the oscillation component $A_\mathrm{large}$. The overall temperature dependence of $A_\mathrm{large}$ in $x$ = 0.05 and 0.1 samples are similar, and the value are comparable to the $x$ = 0 sample at about 20 K. At low temperatures, $A_\mathrm{large}$ in the $x$ = 0 sample is larger than that in the $x$ = 0.05 and 0.1 samples, and the $x$ = 0.03 sample is smaller than the others even at about 20 K. It should be mentioned that due to the pulsed muon with a distribution width of about 70 nano-second, the spectrum with fast muon frequency will be deformed in the early time region and a loss of asymmetry usually appears. Therefore, the extracted amplitude of $A_\mathrm{large}$ should be underestimated, especially at low temperatures where the internal field becomes larger.
\begin{figure}
\includegraphics[width=8.6cm]{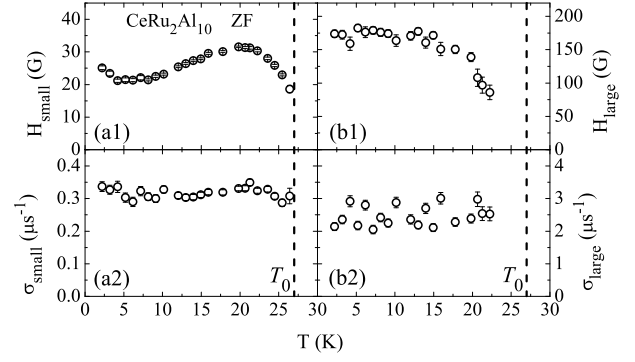}
\caption{(Color online) (a1)(b1) Temperature dependence of the extracted internal fields at the muon sites. (a2)(b2) The damping rates of the muon-spin precessions. \label{undoped_field}}
\end{figure}

Figures \ref{undoped_field}(a1)-\ref{undoped_field}(b2) show the temperature dependence of the extracted internal fields at the muon sites as well as the damping rates for CeRu$_2$Al$_{10}$. As shown in Fig. \ref{undoped_field}(a1) and \ref{undoped_field}(b1), $H_\mathrm{small}$ and $H_\mathrm{large}$ exhibit different temperature dependent behaviors. $H_\mathrm{small}$ begins to increase below $T_0$ as the system becomes long-ranged magnetically ordered, but starts to decrease below about 20 K, which is consistent with previous $\mu$SR results.\cite{Khalyavin_muSR, Kambe_muSR} However, $H_\mathrm{small}$ tends to increase again below about 5 K, which is in contrast to previous study, where $H_\mathrm{small}$ shows a further decrease. On the contrary, $H_\mathrm{large}$ increases below $T_0$ and saturates to the value of about 180 G below about 15 K, as shown in Fig. \ref{undoped_field}(b1). Due to the small amplitude of the oscillation and the fast damping rate, we cannot obtained reliable $H_\mathrm{large}$ just below $T_0$. The temperature dependence of $H_\mathrm{large}$ below 20 K is also different from that in the previous result,\cite{Khalyavin_muSR} which showed decreases behavior below about 20 K, similar to the behavior of $H_\mathrm{small}$. Figures \ref{undoped_field}(a2) and \ref{undoped_field}(b2) show the damping rates $\sigma_\mathrm{small}$ and $\sigma_\mathrm{large}$ of each muon-spin precessions, respectively. These two parameters can reflect the distribution width of the internal field at the muon sites, and exhibit similar temperature dependent behavior as the internal field. A decrease below about 20 K and a slight upturn below about 5 K are observed in the temperature dependence of $\sigma_\mathrm{small}$, while $\sigma_\mathrm{large}$ is almost temperature independent below 20 K.
\begin{figure}
\includegraphics[width=7cm]{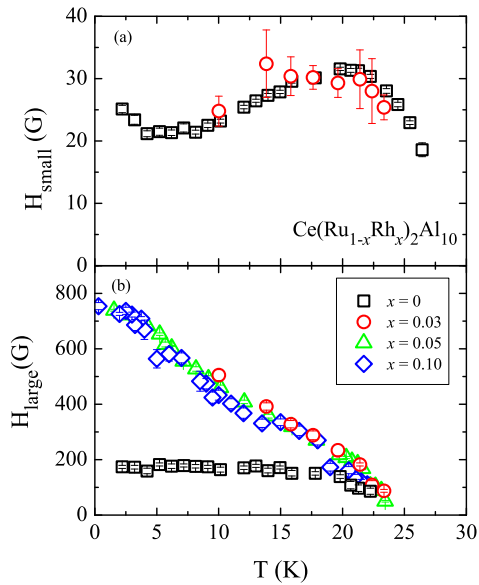}
\caption{(Color online) Temperature dependence of the internal field at the muon sites in Ce(Ru$_{1-x}$Rh$_x$)$_2$Al$_{10}$. (a) The small field component. (b) The large field component.\label{combined_field}}
\end{figure}

Figures \ref{combined_field}(a) and \ref{combined_field}(b) plot the extracted internal fields against temperatures in the $x$ = 0.03, 0.05 and 0.1 samples. For the convenience of comparison, the results obtained from the $x$ = 0 sample are also plotted. In the $x$ = 0.03 compound, muon-spin precession with two frequencies is still observed. As can be seen, $H_\mathrm{small}$ coincides very well with that in the $x$ = 0 compound, but the other component $H_\mathrm{large}$ deviates from that in the $x$ = 0 sample in the measured temperature range. In the $x$ = 0.05 and 0.1 samples, only one frequency is observed, the extracted temperature dependence of the internal fields coincide very well with the $H_\mathrm{large}$ extracted from the $x$ = 0.03 sample, and deviate from the mean-field-like behavior. At the base temperature, $H_\mathrm{large}$ in the Rh-doped sample is about 800 G, which is about four times larger than that in the $x$ = 0 sample.

\section{Discussion}

\begin{figure}
\includegraphics[width=7cm]{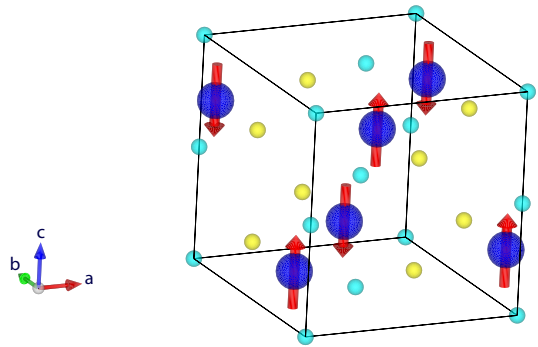}
\caption{(Color online) Unit cell of CeRu$_2$Al$_{10}$ is shown with only Ce ions represented by blue balls. The red arrows indicate the direction of spins. The yellow and cyan balls represent the muons stopped at the (0.5, 0, 0.25) and 4\textit{a} sites proposed by Adroja \textit{et al}.\cite{Adroja_muSR} and Kambe \textit{et al}.\cite{Kambe_muSR}, respectively. \label{CeRu2Al10_structure}}
\end{figure}

Firstly, we discuss the temperature dependence of internal field on the CeRu$_2$Al$_{10}$. The $H_\mathrm{small}$ and $H_\mathrm{large}$ exhibit different temperature dependent behaviors as shown in Fig. \ref{undoped_field}. $H_\mathrm{large}$ shows mean-field-like behavior while $H_\mathrm{small}$ does not. Neutron scattering experiment on both CeRu$_2$Al$_{10}$ and CeOs$_2$Al$_{10}$ single crystals has shown that the Ce$^{3+}$ magnetic moment exhibits mean-field-like ordering and no further magnetic transition is observed below $T_0$.\cite{Kato_neutron_single_crystal} In CeOs$_2$Al$_{10}$, the dip in the temperature dependence of the internal field (first component as described in the paper) is suggested to be probably related to the structural transition and resistivity anomaly.\cite{Adroja_muSR} On the other hand, since the internal field behavior depends on the muon site, we consider that for an antiferromagnet in the ZF condition, the internal field at the muon site can be expressed as $\textbf{H}_\mu=\textbf{H}_{\mathrm{dip}}+\textbf{H}_{c}$, where $\textbf{H}_\mathrm{dip}$ is the dipolar field from the Ce$^{3+}$ moments and $\textbf{H}_{c}$ is the Fermi contact field from the polarized conduction electrons at the muon site. $\textbf{H}_\mathrm{dip}$ is proportional to Ce$^{3+}$ moments and usually show similar temperature dependence, therefore, the decrease of $H_\mathrm{small}$ below about 20 K and the increase again below about 5 K should not originate from $\textbf{H}_\mathrm{dip}$. Alternatively, it can be expected to be from the Fermi contact term $\textbf{H}_c$. Since $\textbf{H}_c$ is closely related to the polarization of conduction electrons at the muon site, we can then expect that a change of Fermi surface or the density of state (DOS) at the Fermi level, $\varepsilon_\mathrm{F}$ may exist below about 20 K. Indeed, from the thermoelectric power measurement on CeRu$_2$Al$_{10}$, anomalies have been observed along the three crystal axes using a single crystal\cite{Tanida_Thermal_power} and two peaks using a polycrystalline sample\cite{Strydom_thermo, Muro_Ru_thermo} at about 20 K and 5 K, respectively. For the CeOs$_2$Al$_{10}$ case, a minimum in the temperature dependence of thermoelectric power was also observed at around 10 K,\cite{Lue_thermo_Os} closed to the dip of the temperature dependence of internal field from $\mu$SR. Although the interpretation of the thermoelectric power is not simple, a change of the slope of DOS at $\varepsilon_\mathrm{F}$ can be related.\cite{Tanida_Thermal_power} Furthermore, from NQR measurement on both $T$ = Ru and Os samples at the Al(5) site,\cite{Matsumura_Os_NQR} the $1/T_1T$ show a small enhancement below about 5 K for the $T$ = Ru sample, and a significant increase below about 10 K for the $T$ = Os sample, indicating that the DOS at $\varepsilon_\mathrm{F}$ is changed, and the temperatures coincide well again with the upturn of internal field from the $\mu$SR results both on the two samples.

From previous NQR measurement on CeRu$_2$Al$_{10}$, the internal fields at the Al sites show mean-field-like behavior and saturate at low temperatures.\cite{Matsumura_reconciliation} The different results between the NQR and $\mu$SR may rely on the different positions of the Al sites and the muon stopping site in the lattice, and the dipolar field dominate the Fermi-contact type hyperfine field at the Al site. The significant influence of the Fermi-contact field for the $H_\mathrm{small}$ component can be due to the small dipolar field at this site and comparable with the Fermi-contact field. Combining the neutron scattering, thermoelectric power, NQR and $\mu$SR results both on CeRu$_2$Al$_{10}$ and CeOs$_2$Al$_{10}$, it is suggested that the anomalous temperature dependence of $H_\mathrm{small}$ observed by $\mu$SR experiment is closely related to the change of conduction electron state and thus may be related to the Fermi-contact type interaction between the muon and conduction electrons.

Next, we discuss the muon stopping sites in the doped samples. Since two muon frequencies are observed on CeRu$_2$Al$_{10}$, two inequivalent muon sites are expected. In a former study on CeRu$_2$Al$_{10}$, Kambe \textit{et al.} suggested the muon site at the 4\textit{a} site between the two nearest Ce ions.\cite{Kambe_muSR} Adroja \textit{et al.} performed $\mu$SR experiment on CeOs$_2$Al$_{10}$ and proposed the muon site (0.5, 0, 0.25).\cite{Adroja_muSR} Both sites are determined by the nuclear dipolar field calculations. The two proposed muon sites 4\textit{a} and  (0.5, 0, 0.25) are depicted as the cyan and yellow balls, respectively, in Fig. \ref{CeRu2Al10_structure}. It can be found that the equivalent positions for the two sites in the unit cell are 4 : 4, which is close to the ratio of $A_\mathrm{small}/A_\mathrm{large} = 0.2 : 0.15$ at the base temperature in the $x$ = 0 sample, as shown in Fig. \ref{normalized_amplitude}. As mentioned above, $A_\mathrm{large}$ is underestimated due to the pulsed muon source, therefore, the ratio can be expected to be more closed to 1 : 1 in the reality. Moreover, since the 4\textit{a} site is just at the center of two nearest neighbor Ce ions, the internal field at this site is expected to be zero in the ideal case, a small internal field may come from a small deviation from this site or the magnetic moments are not ideally antiparallel aligned, as suggested by Kambe \textit{et al}.\cite{Kambe_muSR} Therefore, we propose that the $H_\mathrm{small}$ and $H_\mathrm{large}$ components may originate from the muons stopping at the 4\textit{a} and (0.5, 0, 0.25) sites, respectively, in our present study of CeRu$_2$Al$_{10}$.

In the $x$ = 0.05 and 0.1 samples, only one muon frequency is observed, which seems to indicate that just one muon site appears in the sample. If this is the case, taking into account that the internal field at the base temperature is about 800 G, which is close to the limitation of the RIKEN-RAL facility, the initial asymmetry of the spectrum at the base temperature should be close to 1/3 of the total initial asymmetry since the amplitude of the oscillation will be largely lost due to the pulsed muon structure. By examining the spectra in Fig. \ref{x_005_010_spectrum}(a3) and \ref{x_005_010_spectrum}(b3), the initial asymmetries are about 6\% and 6\%, respectively, at the base temperature for the $x$ = 0.05 and 0.1 samples, which are clearly larger than the expected 1/3 of the total asymmetry of each case. This observation indicates that there may exist another component with much smaller internal field which will not lead to a significant loss of the initial asymmetry. The existence of two muon sites with large and small field components, respectively, is also suggested by the LF measurement as shown in Fig. \ref{x_005_LF}. The recovery of the long time spectrum with the initial asymmetry not affected when LF $<$ 100 G can be attributed to the decoupling of the muon-spin from the small field component. When the LF further increases, the muon spins begin to decouple from the larger internal field and finally the full asymmetry is achieved. Usually, when the external field is about five times larger than the internal field, the resultant vector sum of the total field at the muon site will be nearly parallel to the external field, namely, the muon-spin is nearly decoupled from the internal field, therefore, we can estimate the small internal field from the decoupling behavior in LF = 100 G that it is of the order 100/5 = 20 G at the base temperature, which is close to the value of $H_\mathrm{small}$ obtained in the $x$ = 0 sample. The absence of the muon frequency from the small field component may be due to the large static field distribution which smears out the oscillation due to the substitution of Rh for Ru, which also introduces disorder in the sample. Such a large internal field distribution can also be observed in the $x$ = 0.03 sample, in which two muon frequencies are observed at high temperatures and become unobservable at low temperatures, as shown in Fig. \ref{x_003_spectrum}(d). This may also account for the small amplitudes of the oscillation components in the $x$ = 0.03 sample and the $A_\mathrm{slow}$ becomes larger than the 1/3 value.

Next, we discuss the evolution of internal fields at the muon sites with Rh-doping in Ce(Ru$_{1-x}$Rh$_x$)$_2$Al$_{10}$. Neutron scattering experiments have confirmed that the $m_\mathrm{AF}$ of Ce$^{3+}$ in the $x$ = 0 sample is along the \textit{c}-axis with a value of 0.34-0.42 $\mu_\mathrm{B}$/Ce.\cite{Khalyavin_muSR, Kato_neutron_single_crystal} Magnetic susceptibility measurement on $x$ = 0.05 sample\cite{Kondo_Rh_doped} implies that the $m_\mathrm{AF}$ is along the \textit{a}-axis and the \textit{c-f} hybridization decreases, hence the 4\textit{f} electrons are more localized. As shown in Fig. \ref{combined_field}(b), the $H_\mathrm{large}$ in the $x$ = 0 sample is about 180 G at low temperatures, while it is about 800 G in the Rh-doped samples. The enhancement may come from the change of magnetic structure and/or the enhanced magnetic moment of Ce$^{3+}$ due to the decrease of \textit{c-f} hybridization. In order to manifest the influence of each possibility, we do simple dipolar field calculations at the suggested muon site (0.5, 0, 0.25). Since the 4\textit{a} site is just at the center of the two nearest Ce ions, the dipolar field calculation at this site is always zero, this may also account for the small change of $H_\mathrm{small}$ in $x$ = 0.03 sample and the roughly estimated value in the $x$ = 0.05 sample compared with the undoped sample. The lattice parameters are adopted from the result of neutron scattering at 30 K.\cite{Khalyavin_muSR} The magnetic structure is the same as determined by neutron scattering experiment with the $m_\mathrm{AF}$ along the \textit{c}-axis,\cite{Khalyavin_muSR} and then all the spins are simply rotated uniformly by 90$^\circ$ to be along the \textit{a}-axis and \textit{b}-axis, respectively. The $m_\mathrm{AF}$ of 0.42 $\mu_\mathrm{B}$/Ce is taken from the neutron scattering experiment on the CeRu$_2$Al$_{10}$ single crystal.\cite{Kato_neutron_single_crystal} In order to take into account the more localized 4\textit{f} electrons in the Rh-doped samples, we safely calculate the dipolar field with the $m_\mathrm{AF}$ of two times larger than that in the undoped sample. The calculation results are shown in Table \ref{dipolar_field}.

The internal field is enhanced by about 10 times when the Ce$^{3+}$ moments orientate from the \textit{c}-axis to the \textit{a}-axis with the value of Ce$^{3+}$ moments fixed. On the other hand, the increase of $m_\mathrm{AF}$ affects the results less significantly, and the dipolar field is approximately proportional to the increase of $m_\mathrm{AF}$. This simple calculation reveals that the enhancement of internal field can be mainly attributed to the change of magnetic structure. This result is consistent with the proposal raised by the magnetization measurement.\cite{Kondo_Rh_doped} We also note that the absolute value of the internal field is largely deviated from the experimental value suggesting that other interactions beyond the dipolar interaction should be considered.

\begin{table}
\caption{Dipolar field calculation at the muon site (0.5, 0, 0.25) proposed by Adroja \textit{et al}.\cite{Adroja_muSR} $H_i$ ($i = a, b,c$) is the internal field along the $i$ direction. The unit of $H_i$ is Gauss (G), $\mu_\mathrm{B}$ is the Bohr magneton. \label{dipolar_field}}
\begin{ruledtabular}
\begin{tabular}{lc|ccc}
\multicolumn{2}{c|}{}&$H_a$&$H_b$&$H_c$\\
\hline
         &\textit{m}$_\mathrm{\mathrm{AF}}$ $\parallel$ a&142 & 0 &0\\
0.42 $\mu_\mathrm{B}$/Ce & \textit{m}$_\mathrm{\mathrm{AF}}$ $\parallel$ b& 0 & 155 &0\\
         & \textit{m}$_\mathrm{\mathrm{AF}}$ $\parallel$ c&0&0&13\\
\hline
         & \textit{m}$_\mathrm{\mathrm{AF}}$ $\parallel$ a& 286 & 0 &0\\
0.85 $\mu_\mathrm{B}$/Ce& \textit{m}$_\mathrm{\mathrm{AF}}$ $\parallel$ b& 0 & 313 &0\\
         & \textit{m}$_\mathrm{\mathrm{AF}}$ $\parallel$ c& 0 & 0 &26\\
\end{tabular}
\end{ruledtabular}
\end{table}

Although the change of magnetic structure can partially explain the enhancement of internal field, it still cannot account for the change of temperature dependence of $H_\mathrm{large}$, as shown in Fig. \ref{combined_field}(b). $H_\mathrm{large}$ exhibits mean-field-like increase below $T_0$, and tends to saturated to about 180 G in the $x$ = 0 sample, while it is almost proportional to the temperature in the Rh-doped samples, clearly deviates from the mean-field-like behavior. Moreover, $H_\mathrm{large}$ in the Rh-doped samples follow a universal curve, implying that they should have the same origin. The non-mean-field like behavior of $H_\mathrm{large}$ in Rh-doped samples may be caused by the intrinsic development of the magnetic structure or by the hyperfine interaction between muon and conduction electrons as discussed for the $H_\mathrm{small}$ behavior in the $x$ = 0 compound. In order to clarify these effects, detailed numerical calculation of the hyperfine field and neutron scattering experiment on the Rh-doped samples are highly desirable.

By examining the $x$-dependence of the internal field behavior, it is found that the magnetic ground state of Ce(Ru$_{1-x}$Rh$_x$)$_2$Al$_{10}$ is drastically changed from $x$ = 0 to $x$ = 0.1, and the boundary should locate around $x$ = 0.03. The substitution of Rh for Ru introduces one more electron and can be regarded as electron doping, which can change the $\varepsilon_\mathrm{F}$ and thus the Fermi vector $k_\mathrm{F}$. As a result, the \textit{c-f} hybridization will be tuned as well, as reveal by the magnetization measurement which show more localized 4\textit{f} electrons.\cite{Kondo_Rh_doped} However, the doping level is quite small, and the change of $k_\mathrm{F}$ is not expected to be pronounced. Therefore, the drastic change of the magnetic ground state by such a small perturbation is unusual. We note that by applying magnetic field along the \textit{c}-axis in CeRu$_2$Al$_{10}$, $m_\mathrm{AF}$ orientates from the \textit{c}-axis to the \textit{b}-axis, which is also unusual because the \textit{b}-axis is the hardest axis.\cite{Tanida_fine_structure, Tanida_NMR} Tanida \textit{et al.} speculate that the strong \textit{c-f} hybridization  along the \textit{a}-axis prevents the $m_\mathrm{AF}$ to be along the \textit{a}-axis as proposed by Kondo \textit{et al}.\cite{Kondo_high_field_Os} In our study of Rh substitution, the \textit{c-f} hybridization becomes weaker and $m_\mathrm{AF} \parallel a$ can be realized.  Recently, Khalyavin \textit{et al}. reported the neutron scattering results on the CeOs$_{1.84}$Ir$_{0.16}$Al$_{10}$ system, which has similar electron-doping effect,  and showed that the $m_\mathrm{AF}$ is found to be along the \textit{a}-axis.\cite{Khalyavin_OsIr_neutron} These results suggest that the magnetic structure in the undoped sample is fragile and the strength of \textit{c-f} hybridization may reside in the critical region, a small perturbation such as chemical doping will then easily tune the system to be in different ground states.

\section{Summary}

We have performed $\mu$SR experiments on the Rh-doped Ce(Ru$_{1-x}$Rh$_x$)$_2$Al$_{10}$ with $x$ = 0, 0.03, 0.05 and 0.1. Two muon frequencies are observed in the $x$ = 0 sample, while only one is present in the $x$ = 0.05 and 0.1 samples, which is attributed to the broad static field distribution at the muon site and thus the muon-spin precession is smeared out. The internal field at the corresponding muon site is enhanced by about four times by Rh-doping, which supports the $m_\mathrm{AF}$ being along the \textit{a}-axis in the Rh-doped sample as revealed by our simple dipolar field calculation. The boundary of different magnetic ground states is identified around $x$ = 0.03, suggesting that the magnetic structure of CeRu$_2$Al$_{10}$ is not so robust and can be easily tuned by external perturbations such as chemical doping. We also propose that the anomalous temperature dependence of $H_\mathrm{small}$ in CeRu$_2$Al$_{10}$ is attributed to the Fermi-contact type hyperfine interaction between muons and conduction electrons by comparing previous neutron scattering, thermoelectric power, NQR and $\mu$SR experimental results with the current study. The non-mean-field-like behavior of $H_\mathrm{large}$ in the Rh-doped samples is still not well understood, and future study by neutron scattering experiment and hyperfine field calculation are highly desirable.

\bibliography{CeRu2Al10}

\end{document}